\documentclass{article}
\usepackage{spconf,amsmath,graphicx}
\usepackage[section]{placeins}
\usepackage{url}
\usepackage{relsize}
\usepackage{hyperref}
\usepackage{epstopdf}
\usepackage{enumerate}
\usepackage{algorithmicx}
\usepackage{algorithm}
\usepackage{amsmath}
\usepackage[noend]{algpseudocode}
\usepackage{float}
\usepackage{color}
\usepackage{makeidx}
\usepackage{bbm}
\usepackage{graphicx}
\usepackage{lipsum}
\usepackage{subcaption}
\usepackage[section]{placeins}

\usepackage{setspace}
\usepackage{lipsum}
\usepackage{adjustbox}
\usepackage{colortbl}
\usepackage{hhline}

\usepackage{multicol}
\definecolor{Gray}{gray}{0.88}
\definecolor{Grayy}{gray}{0.77}
\definecolor{Grayyy}{gray}{0.82}

\usepackage{lipsum}
\let\OLDthebibliography\thebibliography
\renewcommand\thebibliography[1]{
  \OLDthebibliography{#1}
  \setlength{\parskip}{0pt}
  \setlength{\itemsep}{0pt plus 0.3ex}
}

\title{The First Pathloss Radio Map Prediction Challenge}
\name{\c{C}a\u{g}kan Yapar$^{\dagger }$ \quad Fabian Jaensch$^{\dagger }$ \quad Ron Levie$^{\ddagger}$ \quad  Gitta Kutyniok$^{*\S\P}$ \quad Giuseppe Caire$^{\dagger}$ 
}

\address{
    $^{\dagger}$Technical University of Berlin, $^{\ddagger}$Technion – Israel Institute of Technology\\
    $^{*}$Ludwig Maximilian University of Munich, $^{\S}$University of Tromsø \\
$^{\P}$Munich Center for Machine Learning (MCML)
}
\newcounter{CagkanCounter}

\begin{document}
%
\maketitle
\begin{abstract}

To foster research and facilitate fair comparisons among recently proposed pathloss radio map prediction methods, we have launched the  ICASSP 2023 First Pathloss Radio Map Prediction Challenge. 
In this short overview paper, we briefly describe the pathloss prediction problem, the provided datasets, the challenge task and the challenge evaluation methodology. Finally, we present the results of the challenge.

\end{abstract}
%
%
\vspace{-4.2mm}
\section{Introduction} \label{sec:Intro}
\vspace{-2.2mm}
	In wireless communications, the \emph{pathloss} (or \emph{large scale fading coefficient}) quantifies the loss of signal strength between a transmitter (Tx) and a receiver (Rx) due to large scale effects, such as free-space propagation loss, and interactions of the radio waves with the obstacles (which block line-of-sight, for instance buildings, vehicles, pedestrians in urban environments), e.g. penetrations, reflections and diffractions.
	
    Many present or envisioned applications in wireless communications explicitly rely on the knowledge of the pathloss function, and thus, estimating pathloss is a crucial task. Some example use cases include: User-cell site association, fingerprint-based localization, physical-layer security, optimal power control,  path planning, activity detection \cite{RadioUNetTWC}.

    Deterministic simulation methods such as ray-tracing are well-known to provide very good estimations of pathloss values. However, their high computational complexity renders them unsuitable for most of the envisioned applications.

    In recent years, many research groups have developed deep learning-based methods that achieve comparable accuracy to ray-tracing, but with orders of magnitude lower computation time, providing the fast and reliable pathloss estimation required by the applications.


\vspace{-3.2mm}
\section{Datasets}\label{sec:Datasets}
\vspace{-2.2mm}
\subsection{Training Dataset}\label{sec:3DDataset}
\vspace{-2.2mm}
For the training, we provided the challenge participants with the \emph{RadioMap3DSeer Dataset}, which we  set publicly available  as a part of a collection of radio map datasets that we generated under various settings \cite{DataPort2}.

The pathloss radio maps of the dataset were generated based on the simulations by the ray-tracing software \emph{WinProp from Altair} \cite{WinPropFEKO}, on a dataset of urban environments. The city maps were fetched from \emph{OpenStreetMap} \cite{OpenStreetMap} in the cities Ankara, Berlin, Glasgow, Ljubljana, London, and Tel Aviv, amounting to 701 city maps of size $256 \times 256$ meters. All simulations were run with a resolution of 1 meter and saved as images of $256 \times 256$ pixels.
80 rooftop transmitter locations per map were considered, resulting in a total of 56080 simulations. Pathloss values were calculated at 1.5 m from the ground.

The simulations are based on the \emph{Intelligent Ray Tracing (IRT)} \cite{IRT} method.
 For simplicity, all buildings were assumed to have the same generic material property.

 The pathloss values obtained from the simulations were truncated  below a minimum pathloss value and the range between this minimum pathloss value and the maximum pathloss value over the whole simulations was scaled to gray levels between 0 and 1, to save the pathloss radio map simulations as images. The details on the determination of the pathloss truncation value and the applied scaling can be found in \cite{DatasetPaper,RadioUNetTWC} and
 more detailed descriptions of the dataset in \cite{DatasetPaper,DataPort2}.

 Our previously published work \cite{RadioUNetTWC} and its publicly available code were available for the participants as a baseline.

\vspace{-3.2mm}
\subsection{Test Dataset}
\vspace{-2.2mm}

For the evaluation of the participants' methods, we prepared a test dataset which was not published before. 84 city maps of size 256$\times$256 were obtained from \emph{OpenStreetMap} \cite{OpenStreetMap} in Istanbul, resulting in 6720 simulations.
The same dataset generation procedure and simulation parameters were used as for \emph{RadioMap3DSeer} \cite{DatasetPaper,DataPort2}.

\section{The Challenge Task}
\vspace{-2.2mm}

\begin{figure*}[hbt!]
    \centering
    \begin{subfigure}[b]{0.19\textwidth}
        \includegraphics[width=.94\textwidth]{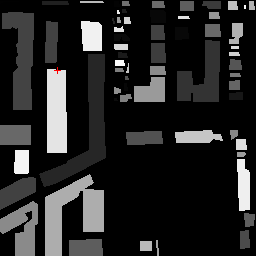}
        \caption{City map w/ height}
        \label{fig:IRT23DGT}
    \end{subfigure}
    \begin{subfigure}[b]{0.19\textwidth}
        \includegraphics[width=.94\textwidth]{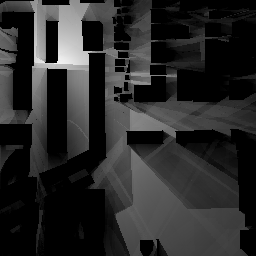}
        \caption{Ground truth}
        \label{fig:IRT23DNaive}
    \end{subfigure}
    \begin{subfigure}[b]{0.19\textwidth}
        \includegraphics[width=.94\textwidth]{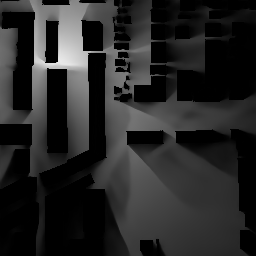}
        \caption{PMNet \cite{PMNet}}
        \label{fig:IRT23DHeight}
    \end{subfigure}  
    \begin{subfigure}[b]{0.19\textwidth}
        \includegraphics[width=.94\textwidth]{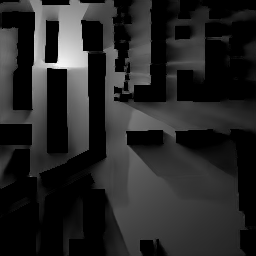}
        \caption{Agile \cite{Agile}}
        \label{fig:IRT23DHeight}
    \end{subfigure}   
    \begin{subfigure}[b]{0.19\textwidth}
        \includegraphics[width=.94\textwidth]{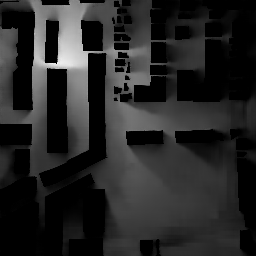}
        \caption{PPNet \cite{PPNet}}
        \label{fig:IRT23DHeight}
    \end{subfigure}
    \caption{Examples from the test set. \textbf{From left to right:} \textbf{a)} Height-encoded city map (the brighter the pixel, the higher is the building. Tx location is shown with a red plus sign), \textbf{b)} Ground truth simulation, \textbf{c,d,e)} Predictions of the successful methods}
    \label{fig:Results7_73}
\end{figure*}

The task of the challenge was to predict the pathloss radio map given the city map and the transmitter location, i.e., the same task and input setting of  deterministic simulation methods like ray-tracing.

The participants were allowed to design their input features (i.e. pre-processing) freely,
as long as the test run-time of the proposed method was orders of magnitude lower than the pathloss simulation by the ray-tracing software.

\vspace{-4.2mm}
\section{Challenge Results}
\vspace{-2.2mm}
\subsection{Evaluation Methodology}
\vspace{-2.2mm}
The participants were asked to submit their radio map predictions for the challenge test set (which was sent them without the ground truth)  along with the code that runs the evaluation. 

While evaluating the prediction performance of the submitted methods, we first set the pixels of the radio map predictions known to be occupied by the buildings to zero, i.e., given that the ground truth value at such pixels is zero, the prediction error was guaranteed to be zero for such pixels.

 We evaluated the accuracy by the root mean square error $\textup{RMSE} =  \sqrt{\frac{1}{|\mathcal{T}|}\sum_{n \in \mathcal{T}} \textup{RMSE} (n)^2}$ where $\mathcal{T}$ is the test set and $\textup{RMSE} (n)$ is the RMSE for the radio map $n$, defined as
\vspace{-2.2mm}
\begin{equation}\label{eq:RMSE}
    \textup{RMSE} (n)= \sqrt{\frac{1}{RC}\ \sum_{i=1}^{R} \sum_{j=1}^{C}\left(\tilde{\textup{P}}_{\textup{L}}^{(n)}(i,j)-\textup{P}_{\textup{L}}^{(n)}(i,j)\right)^2}
\end{equation}
where $\tilde{\textup{P}}_{\textup{L}}^{(n)}$ and $\textup{P}_{\textup{L}}^{(n)}$ are the predicted and the ground truth radio maps, $R$ and $C$ are the number of rows and columns in a radio map image ($R=C=256$ in our dataset), respectively. 
\vspace{-6.2mm}
\subsection{Evaluation Results}
\vspace{-1.2mm}
\begin{table}[!h]
\vspace{-2.2mm}
		\caption{\small Accuracies of the submitted methods on the test set} \label{table:RMSETable}
	\renewcommand{\arraystretch}{1}

	\centering
	\scalebox{0.95}{
	\begin{tabular}{c|c}
		\hline	
		\rowcolor{Grayy}  {\bfseries  Method }& \bfseries RMSE\\
		\hline
	
		PPNet \cite{PPNet}  & $0.0507$\\
            \hline	
		\hline		
		Agile  (MSE) \cite{Agile} & $0.0514$\\
	
		\hline
             Agile  (MSE, LoS) \cite{Agile} & $0.0461$\\
	
		\hline
              Agile  (KL, LoS) \cite{Agile} & $0.0451$\\
	
		\hline
            \hline
             
		PMNet (w/ Fine Tuning) \cite{PMNet} & $0.0959$\\
             \hline

             	PMNet (w/ Data Aug.) \cite{PMNet}& $0.0633$\\
             \hline
		PMNet ($\frac{H}{8}\times\frac{W}{8}$) \cite{PMNet}& $\mathbf{0.0383}$\\

		\hline

	\end{tabular}
	}

\end{table}
We summarize the accuracies of the submitted methods in Table \ref{table:RMSETable} and show the prediction results for a sample from the test set in Fig. \ref{fig:Results7_73}. Based on our evaluations and the declarations of the participants, a large degradation of performance of all the submitted methods on the test set is observed (with respect to testing on a hold-out subset of \emph{RadioMap3DSeer}).

All participants reported run-times of about $\sim$10 ms.

We would like to note here that the RMSE calculations reported in \cite{PMNet} differ  from the one explained here \eqref{eq:RMSE} and in \cite{PPNet,Agile}.
The RMSE results presented in \cite{PMNet} were found by evaluating on a hold-out set of \emph{RadioMap3DSeer} and seemingly by averaging of RMSEs calculated on mini-batches of size 16, and without setting the building pixels to zero.
Also, we couldn't verify PMNet (w/ Fine Tuning)  version to yield the given results in \cite{PMNet}, as we observed worse performance also in the hold-out \emph{RadioMap3DSeer} subset they apparently  used.

 Nevertheless, the best performing method (PMNet ($\frac{H}{8}\times\frac{W}{8}$)) of \cite{PMNet} demonstrated a remarkable performance on the challenge test set, ranking the first among all the submitted methods.

\bibliographystyle{IEEEbib}
 
{
\footnotesize
\bibliography{pub}
}

\end{document}